\documentclass[11pt]{llncs}

\usepackage{mathpartir}
\usepackage{amsmath,amssymb}
\usepackage{xcolor}
\usepackage{stmaryrd}
\usepackage{mathbbol}
\usepackage{mathtools}
\usepackage{graphicx}
\usepackage{upgreek}
\usepackage{tikz}
\usepackage{tikzscale}
\usepackage{url}
\usepackage{mathabx,stmaryrd}
\usepackage{fancyvrb}
\usepackage{a4wide}
\usepackage[mathscr]{euscript}

\usepackage{enumitem}
\begin{document}

\title{Exploring Moral Exercises for Human Oversight of AI systems: Insights from Three Pilot Studies}

\author{Silvia Crafa\inst{1} \and Teresa Scantamburlo\inst{2}}

\institute{University of Padova \and Ca' Foscari University of Venice}

\maketitle

\begin{abstract}
 This paper elaborates on the concept of moral exercises as a means to help AI actors cultivate virtues that enable effective human oversight of AI systems. We explore the conceptual framework and significance of moral exercises, situating them within the contexts of philosophical discourse, ancient practices, and contemporary AI ethics scholarship.
We outline the core pillars of the moral exercises methodology — eliciting an engaged personal disposition, fostering relational understanding, and cultivating technomoral wisdom — and emphasize their relevance to key activities and competencies essential for human oversight of AI systems. Our argument is supported by findings from three pilot studies involving a company, a multidisciplinary team of AI researchers, and higher education students. These studies allow us to explore both the potential and the limitations of moral exercises. Based on the collected data, we offer insights into how moral exercises can foster a responsible AI culture within organizations, and suggest directions for future research.
\end{abstract}
\keywords{Moral exercises, human oversight, human agency, moral cultivation, virtues}

\section{Introduction}

Ensuring human oversight of artificial intelligence (AI) systems is a fundamental requirement for trustworthy AI. However, existing approaches to oversight often focus on procedural safeguards and compliance measures, overlooking the deeper ethical dimensions necessary for meaningful human involvement. This paper introduces moral exercises as a novel approach to enhancing the ethical reasoning and judgment capabilities of AI actors. Moral exercises advance the perspective that human oversight is not merely a procedural or technical function but, instead, involves moral cultivation and ethical discernment in techno-social contexts.\cite{vallor2016technology,rulesDaston}. 

Our study provides three contributions. First, we develop the framework of \emph{moral exercises}, resting on philosophical traditions and contemporary AI ethics scholarship 
that emphasizes the need for moral cultivation beyond formal governance mechanisms.
We higlight that they are 
not intended to instruct what is an ethical action, but to
develop ethical guidance in collective discernment. In this sense they are process-oriented rather than solution-driven.

Second, we outline the methodological pillars of moral exercises --personal commitment,  relational understanding, and technomoral wisdom— showing how they can be structured to engage participants in ethical reflection within  concrete AI contexts. Third, we assess the feasibility and impact of moral exercises  through three pilot studies conducted in different socio-technical contexts: a business setting, an academic research workshop, and a university course on AI ethics. These studies provide empirical insights into the potential of moral exercises to foster engagement, enhance ethical awareness, and strengthen professional judgment.

The paper is structured as follows: Section~\ref{sec:moralEx} situates moral exercises within the broader context of human oversight of AI, discussing their theoretical foundations and ethical significance. Section~\ref{sec:pilot} presents the methodology and findings of the three pilot studies, illustrating how moral exercises can be applied in practice. Section~\ref{sec:discussion} offers a critical discussion of the key insights gained from these studies, highlighting both strengths and challenges.
Finally, Section~\ref{sec:discussion} concludes summarizing our results and outlining directions for future research.

By advancing moral exercises as a structured yet flexible approach to ethical reasoning that aims to be transformaive rather than informative, we intend to contribute to the ongoing discourse on responsible AI, offering a practical tool to enhance human agency in AI oversight.

\section{MORAL EXERCISES IN THE CONTEXT OF HUMAN OVERSIGHT}
\label{sec:moralEx}
Human agency and oversight are key requirements for trustworthy Artificial Intelligence (AI) \cite{hlegTAI}, shared across various AI policies \cite{jobin2019global}, particularly in the context of algorithmic decision-making \cite{peeters2020agency}. Broadly speaking, they refer to the central role of the human being throughout an AI life cycle, also known as `human centrism' \cite{enqvist2023human}, ensuring that people remain in control of, and accountable for, AI systems so that they  do not lead to harmful or undesirable outcomes. 
To date, much of the discussion has focused on human oversight, commonly understood as the active supervision of AI systems by individuals. In practice, this requirement can take many forms, such as governance frameworks \cite{hlegTAI}, procedures for monitoring system performance, and mechanisms for addressing anomalies \cite{EUaiAct}. For instance, the European AI Act emphasizes that those responsible for oversight must have the necessary skills, training, and authority to effectively perform their duties \cite{EUaiAct}. However, defining what this entails is a contentious issue, presenting both conceptual and practical challenges.

\vspace{-0.2cm}
\subsection{The challenges of human oversight}

AI ethics scholarship has explored human oversight from different angles. Researchers have examined the concept in policy frameworks \cite{koulu2020proceduralizing,green2022flaws} and recent legislation \cite{laux2023institutionalised,enqvist2023human,beck2024human}. They have investigated essential prerequisites for its effective application \cite{kyriakou2023humans,methnani2021let,sterz2024quest} demonstrating that mere human presence is not sufficient.
Researchers have also pointed out ``a control problem'' described as the tendency of human overseer to exhibit over-reliance or excessive skepticism \cite{zerilli2019algorithmic}. Mixed evidence confirmed this issue.
There are studies showing that humans may become overly dependent on automated system outputs and, as a consequence,  commit error or omissions \cite{parasuraman2010complacency}. While others suggest that they can be reluctant to trust AI decisions \cite{longoni2019resistance,mahmud2022influences}, and manifest what has been called ``algorithmic aversion'', a bias toward human forecasts over more accurate algorithmic predictions \cite{dietvorst2015algorithm}, and struggle to assess algorithmic performance \cite{green2019principles,poursabzi2021manipulating}. 
This literature suggests that human oversight comprises multiple capabilities and skills. These include technical competences (e.g. related to AI system's capabilities and limitations \cite{EUaiAct}), but also pragmatic and organizational skills. For instance, it is important to consider the allocation of tasks between humans and AI systems, as well as protocols of action in case of unexpected systems' behavior. Interestingly, technical and organizational dimensions in oversight efforts are not new in the history of automation and partly echo the roles of human supervisors and operators who worked alongside calculating machines in mid-nineteenth-century astronomical observatories and census bureaus \cite{rulesDaston}.

However, human oversight extends beyond the mere technical and organizational control of an AI system. As well as epistemic and governance capabilities, Sterz and colleagues suggest more fundamental qualities such as self-control and the willingness to fit the overseeing role \cite{sterz2024quest}. The latter closely relate to moral deliberation and human responsibility, which are far less investigated in human oversight literature compared to technical and organizational dimensions. Instead, most works approach human oversight primarily through procedural safeguards, aligning with the `algorithmisation' of decision-making processes across bureaucracies and institutional contexts \cite{koulu2020proceduralizing}. This procedural focus creates a tension: although oversight is meant to introduce discretion and human autonomy in determining the `best' application of an algorithmic output, it is paradoxically framed according to the same logic as algorithms — treated as an encoded procedure for solving problems rather than as an exercise of moral judgment. 

The framework of moral exercise that we propose aims to address this friction by positioning the human being at the heart of navigating the complexities and ethical dimensions of AI systems. It grounds human oversight in fundamental dispositions, requiring personal engagement with critical questions, active listening and the balancing of diverse or even conflicting perspectives. Built upon the assumption that responsibility ``holds also an unavoidable element of personal commitment.'' \cite{gorgoni2021responsibility}, the framework aims at providing concrete proposals to cultivate moral character through relational understanding, reflective self-examination and prudential judgment. 
\vspace{-0.2cm}
\subsection{Moral exercises for ethical reasoning and deliberation}

The framework of moral exercises results from a reaction to tempting directions within AI ethics: the appeal to abstract principles lacking practical guidance, also known as principalism \cite{mittelstadt2019principles}, and the narrow focus on operationalization of values and principles \cite{hagendorff2022blind}. To complement these perspectives it suggests a methodological approach to frame activities that aim to reconnect responsible AI to the role of the acting subject \cite{gorgoni2021responsibility}. 

A moral exercise is a structured activity designed to cultivate and enhance the ethical skills and moral judgment of individuals involved in decision-making processes, particularly in the oversight of algorithmic systems. It aims to develop critical ethical dispositions such as a sense of shared values, prudence, and the ability to evaluate actions against universal moral principles. These exercises are not abstract nor rigid procedures, but flexible tools meant to encourage reflection, dialogue, and responsibility, starting from a concrete scenario involving AI systems. 

The concept of moral exercises is inspired by the decision-making practices of early monastic communities in the first millennium. These communities engaged in discernment activities to guide both individual and collective decisions, emphasizing attentive listening and active engagement with both internal (thoughts, emotions, values) and external (events, interactions, contexts) realities. 
In addition, these small, diverse groups developed collective decisions practices that necessarily had to balance cohesion with different perspectives, address biases and hierarchical structure. 
When adapted to modern techno-social systems \cite{grandi2021discernimento,Sca2023}, these practices provide valuable insights for developing activities that cultivate key ethical dispositions, such as attentiveness to inner speech for distinguishing beliefs and intentions, and sensitivity to others' perspectives.
\vspace{-0.2cm}
\subsection{The pillars of moral exercises}
\label{sec:pillars}

Moral exercises encompass distinct dimensions that do not prescribe a fixed methodology but instead provide guidance for designing activities in specific socio-technical contexts. The following criteria outline key considerations that may serve the implementation of moral exercises in alignment with their inspiring principles (see the discussion in Section~\ref{sec:discussion}).

\begin{itemize}[leftmargin=*]
\item {\bf Personal commitment and self-examination.} 
A key aspect of moral exercises is fostering the appropriate personal disposition of participants, which is essential for the discernment process. Participants are encouraged to activate their moral capabilities and civic virtues. Engaging participants requires drawing on their professional experience and personal judgment skills extending professionalism beyond technical competence \cite{pasquale,moor}. This emphasizes shared human conditions, such as citizenship and collective responsibility, over domain-specific expertise. 
This aligns with the idea that responsibility is essentially a disposition of the agent, since it implies ``a willingness to understand and confront the other’s commitments and concerns with ours, to look for a possible terrain of sharing. It entails readiness to rethink our own problem definition, goals, strategies, and identity.'' \cite{gorgoni2021responsibility}, 
In practice, a moral exercise should actively engage participants in reflecting on and critically engaging with specific AI scenarios. It requires deep listening to one's inner speech \cite{FantiRovetta23} as a source of moral action, while also enacting broad moral capabilities, including emotional and social intelligence \cite{vallor2016technology}.


\item {\bf Relational understanding and group consultation.} In addition to self-reflection, a second pillar of moral exercises recognises the relational nature of ethical reasoning in AI supervision. Ethics scholarship recognizes that ``moral agent is not the person who most successfully detaches her deliberations from her own relational context, rather the one who understands and responds to that context most fully.''\cite{vallor2016technology}. 
Discourse ethics \cite{discourse,test-driven-ethics}
calls for effective communication between  those affected by decisions and proposals, including developers, managers, employees, shareholders, customers, users, and others stakeholders. This approach seeks to integrate diverse ethical perspectives, continuously evaluate alternative actions, and foster constructive, deliberative discussions. However, achieving meaningful dialogue and deliberation is challenging and cannot be reduced to a purely procedural process.

A central element in collective decision-making is ensuring that discussions remain clear, focused, and concise. To this aim, our methodology relies on active group listening and an effective time management that guarantees participation, inclusion, and balanced attention to different voices. By encouraging participants to prepare concise written contributions, this approach minimizes imbalances in participation and helps clarify the group’s collective orientation.
Moral exercises emphasize the importance of listening to diverse voices and creating meaningful conversations among stakeholders \cite{robinson2022voices}. Group listening provides an opportunity to transform individual insights into shared knowledge, encouraging deeper reflection. They help clarify the nuances of broad ethical concepts such as privacy or autonomy, reveal points of convergence and divergence among AI actors, and facilitate a shared understanding of the values at stake in a given AI scenario.

A typical difficulty in discourse ethics is that the discussion  need to accommodate both universal rules and particular situations. Participants must accept that some ethical questions may well not generate universal, but only local, agreement and yet can still be the subject of rational choice \cite{discourse}. The group sharing methodology of moral exercises fosters the the construction of common understanding and then agreement about shared actions.

\item {\bf Structured but flexible methodology, centered on people and technomoral wisdom.} 
Moral exercises integrate seeks to integrate the technical and moral dimensions of human oversight by applying humanist and philosophical traditions to navigate the practical and productive demands of the technological domain. While organizations require structured procedures, translating ethical principles into concrete situations remains challenging.
To bridge this gap, moral exercises use structured and guided conversations, at both individual and group levels, focused on real-world ethical concerns. 

The exercise format, provided through a participant worksheet, organizes the flow of time and activities to ensure meaningful listening and exchange within a community. At the same time, the methodology remains flexible, accommodating different ethical questions — such as identifying shared values or evaluating specific actions — and adapting to various AI stakeholders and system development stages, including training and AI literacy. In Section~\ref{sec:pilot}, we outline the structure of a moral exercise tested in three different contexts, while other blueprints are discussed in \cite{Sca2023}.

At the core of this methodology is the activation of 
\emph{practical wisdom} (or  \emph{technomoral wisdom}), a virtue that integrates intellectual, cognitive, perceptual, emotional, and social capacities to apply moral principles effectively in specific situations \cite{vallor2016technology}. Incorporating this classical virtue into ethical deliberation within techno-social contexts engages a uniquely human ability: the capacity to synthesize logical, technical, moral, and emotional dimensions into an intelligent and context-sensitive response. This process is not a mere synthesis or algorithmic computation but a form of human judgment—the ability to reflect on experience and navigate ethical complexities beyond mere data processing.

%


\end{itemize}


\vspace{-0.2cm}
\subsection{Related ethical frameworks and approaches}
Moral exercises are grounded in classical philosophical traditions that emphasize the anthropological dimensions of moral decision-making \cite{Sca2023}. In the following we explore two contemporary references that shed light on the perspective and orientation of moral exercises.
\label{sec:framework}
\vspace{-0.2cm}
\paragraph{Virtue ethics}

The framework of moral exercises  aligns with \emph{virtue ethics} \cite{sep-ethics-virtue}, an ethical approach that emphasizes the development of distinctive human abilities. Unlike utilitarianism, which prioritizes efficiency and optimization, virtue ethics follows the logic of moral good, seeking a balanced synthesis that gives meaning and direction to moral judgment.
Virtues are dispositions that the person must cultivate in herself, and that once cultivated, lead to deliberate, effective, and reasoned choices of the good. The virtuous state emerges gradually from habitual and committed practice and study of right actions. This suggests that moral exercises should be considered as a regular practice to create real impact. 
The pillars of moral exercises 
closely relate to the application of virtue ethics in the techno-social domain \cite{vallor2016technology}, since this prior work suggested the need for reflective self-examination, moral habituation as a transformative process, relational understanding and the technomoral practical wisdom. Similarly, moral exercises are rooted in the intuition that 
``practical wisdom encompasses considerations of universal rationality as well as considerations of an irreducibly contextual, embodied, relational and emotional nature.''\cite{vallor2016technology}. 
Their methodology intends to foster AI actors in the cultivation of all such tacit and embodied moral knowledge, that cannot be captured in explicit and fixed decision procedures.
\vspace{-0.2cm}
\paragraph{Rules as models} 
Another key reference for moral exercises is the shift from viewing rules as fixed algorithms to understanding them as flexible models, as described in \cite{rulesDaston}. Algorithmic rules are precise and rigid, designed to be followed mechanically, as seen in mathematics or bureaucratic procedures. In contrast, model-based rules serve as behavioral guidelines that require interpretation, judgment, and adaptation, similar to moral or social norms.
Daston \cite{rulesDaston} highlights the distinction between emulating and merely imitating a model, advocating for a view of rules as models to be emulated rather than as algorithms to be mechanically executed. This perspective underscores the positive role of discretion in judgment. Emulating a model involves internalizing a general framework and then adapting it to specific situations, much like how a child learns by absorbing patterns from their parents and improvising in new contexts.
In this view, discretion—including the ability to recognize exceptions—is an inherent part of rule application. Distinguishing between cases that differ in subtle but significant ways requires more than analytical reasoning; it involves cognitive and ethical judgment. According to Daston, discretion draws meaning from experience, linking it to prudence and other forms of practical wisdom, as well as from guiding values that shape moral reasoning. In this sense, discretion is both an intellectual and a moral faculty, and is part of professionalism.

\vspace{-0.2cm}

\section{PILOT STUDIES}
\label{sec:pilot}

\begin{figure}
\hspace*{-0.7cm}
\includegraphics[width=15cm]{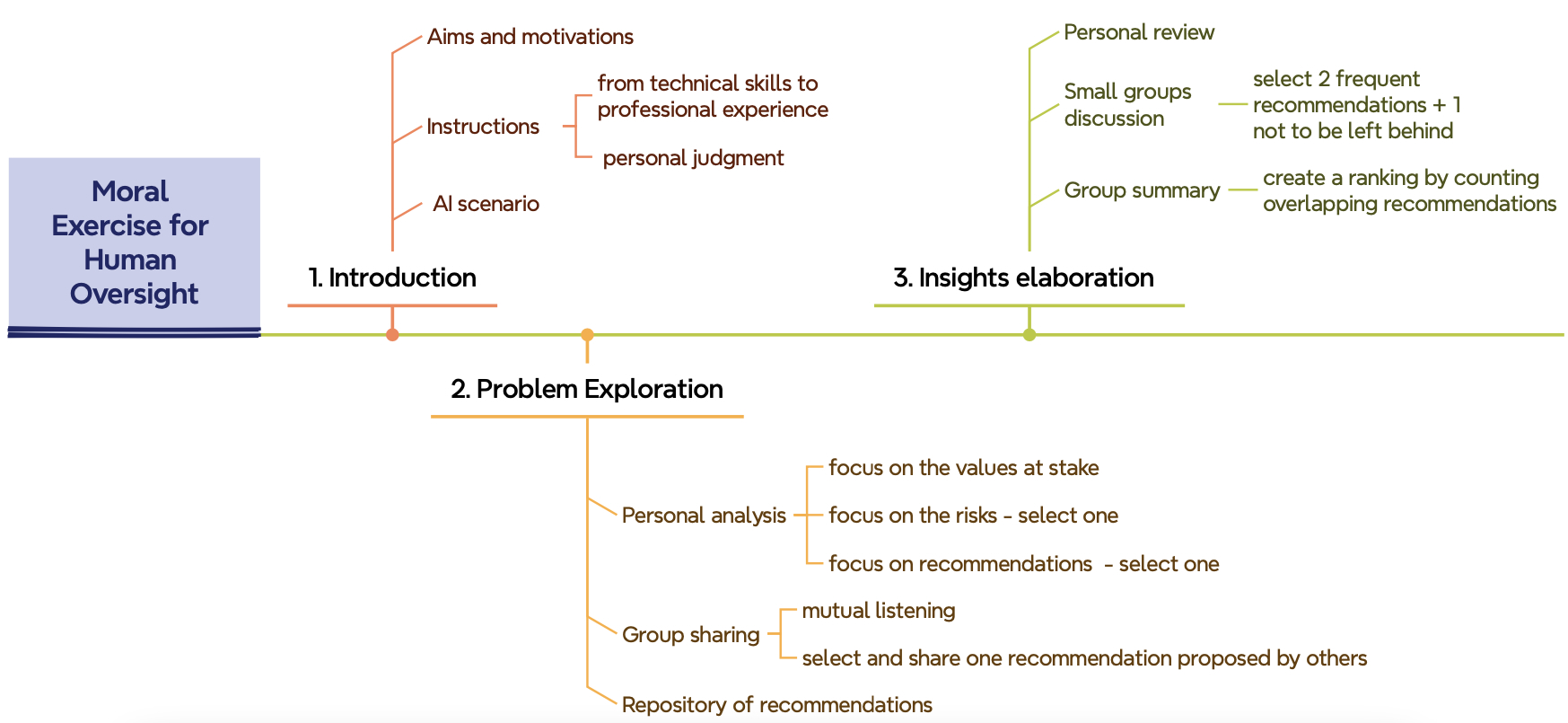}
\caption{Structure of the moral exercise used in pilot studies.}
\label{fig:ex}
\end{figure}



To explore the feasibility and practical implications of moral exercises in a socio-technical context we conducted three pilot studies between June and December 2024. The studies targeted different communities - a company, a group of academic researchers and a group of higher education students. This approach allowed us to capture diverse perspectives and conduct a preliminary assessment of the relevance of moral exercises across different settings. 
Participants were engaged in ethical reflection and conversions through a specific AI application scenario related to the group's experience and context. Each study was conducted in person and moderated by the authors of this work, using task-specific worksheets tailored to the context, the target group and the time constraints. 
\vspace{-0.2cm}
\subsection{Method}

\subsubsection*{Study design} For each pilot study, we gathered preliminary information about the context and participants through online meetings and email exchanges with a contact person from the involved organization. This process allowed us to define a specific purpose for each study, to adapt the activities to the organization's requirements, 
and to carefully select the scenarios in order to enhance participants' engagement.
Worksheets were designed to support participants' activities, drawing on previously developed resources for community discernment \cite{grandi2021discernimento}. The worksheets were divided into sections, each including brief instructions and one or more questions with allocated spaces for participants to take notes. These spaces were intentionally limited (approximately 500–800 characters) to encourage concise responses (see Fig.~\ref{fig:data} and Appendix~\ref{app:sheets}). 

\subsubsection*{Study procedure and process} We structured the moral exercises as a three-step process 
(see Fig.~\ref{fig:ex}). 
For each step, we carefully determined the time allocation to ensure a balanced and effective progression through the activities.
\begin{itemize}[leftmargin=*]
  \item \texttt{Introduction.} Participants were introduced to the aims and underlying motivations of moral exercises. This phase sets the stage 
  by providing information about the necessary dispositions to activate focus attention and group listening (cf. pillars in Section~\ref{sec:pillars}). In particular, participants were encouraged to shift their focus from technical skills to professional experiences and the development of personal judgment capabilities. In addition, participants were provided with a brief description of the AI application scenarios and the associated task, which involved identifying potential risks of the presented application and proposing recommendations to mitigate those risks.
  \item \texttt{Problem exploration.} This phase began with a personal analysis of the AI application. Participants were invited to annotate free thoughts including values at stake, regarding the presented scenarios. They were tasked with annotating potential risks associated with the considered application and selecting one. This task was framed as an exercise in \emph{prioritization rather than synthesis}, guiding participants to focus on a risk they perceived as most critical. This approach was based on the premise that each participant's perspective, while not exhaustive, would serve as a valuable contribution to building a broader understanding of the problem, akin to assembling pieces of a mosaic (cf. pillars in Section~\ref{sec:pillars}). The same process was applied to the formulation of recommendations, with participants asked to explore and prioritize a single recommendation. The activity then moved to a sharing phase, where the insights generated during personal reflection were brought forward for collective consideration. 
  Participants were arranged in a circle, and each person read aloud what they had written during the personal reflection phase, without offering any additional explanation or commentary. While listening, others took notes trying to remain faithful to the insights as originally expressed.  This simple activity required participants to pay close attention to each other, ensuring that speakers felt genuinely heard and that note-takers captured the insights as accurately as possible to their original formulation. After the reading round, each participant were asked to select 
  and read aloud --without adding any comments or further explanation--
  one recommendation from those shared by others that he felt was particularly relevant to the application’s development. 
  The goal was not to summarize or expand on the ideas but simply to highlight the content that stood out most to them. To conclude the sharing phase, participants were asked to submit the recommendation they prioritized from their personal brainstorming, as well as the one they selected from others, to a Wooclap event. This process helped create and visualize a shared repository of recommendations, reflecting priorities at both the individual and group levels. Note that the repository showed overlapping contents but also diverse perspectives since the same recommendation could be expressed differently.
  \item \texttt{Insights elaboration.} This phase aimed at reflecting on the content emerged during the prior step, shaping the repository of collected recommendations into a prioritized ranking. The process began with a brief moment of personal review, allowing participants to pause for considering the ``voices'' that held the most significance for them. Specifically, they were encouraged to identify recommendations that showed convergence (e.g., those frequently highlighted) as well as those representing unique or isolated perspectives that might otherwise be overlooked. The process continued with a brief yet focused discussion in small groups, aimed at identifying a shortlist of recommendations. The groups started by sharing the results of their personal reviews and reported two recommendations with the greatest agreement in a group-specific worksheet. Then the groups were invited to formulate an additional recommendation, which could involve combining similar suggestions or integrating complementary or more original perspectives. This step encouraged creative thinking, allowing for rephrasing and synthesis. The phase concluded with each group sharing their shortlist of recommendations, from which a simple count of overlapping proposals was used to generate a ranked list of priorities.
\end{itemize}

\subsubsection*{Data collection and analysis}
\begin{figure}
\includegraphics[width=14cm]{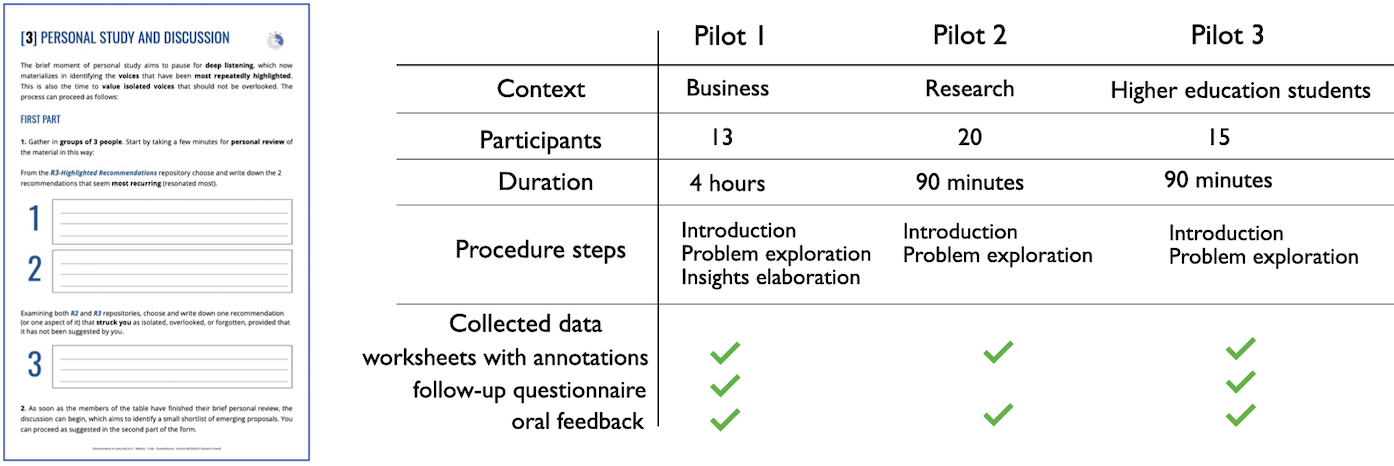}
\vspace*{-0.2cm}
\caption{Sample of worksheet and collected data}
\label{fig:data}
\end{figure}

The overall process generated various types of data (see Fig~\ref{fig:data}), including participants' annotations, Wooclap messages, participants' feedback, and authors' observations. Participants' annotations were collected anonymously through worksheets, 
which were submitted on a voluntary basis. 
At the end of the sessions, an anonymous questionnaire was distributed to gather feedback on the activities. The questionnaire included 5-point scale items to assess participants' experiences (e.g., alignment with expectations, levels of comfort, and perceived usefulness), as well as open-ended questions about aspects they appreciated, potential difficulties, suggested changes, and any additional comments. All data collection was conducted in compliance with the Ethics Review Board and the Data Protection Office of our university. 
The analysis focused on the qualitative datasets; 
the primary aim was to examine the themes that emerged during the exercises, as well as participants' attitudes and behaviors. The analysis assessed the overall effectiveness of the process, identifying both strengths and weaknesses (see Section~\ref{sec:discussion}). A thematic analysis \cite{braun2006using} was conducted on the datasets, with emergent issues (e.g., ambiguous responses or interpreting hand-written text) discussed in online and in-person meetings of the autohrs (who were also the exercises' moderators). 
In each pilot study, the outcomes were categorized into three types: those related to the scenario, those reflecting participant perspectives, and those associated with the exercises.

\vspace{-0.2cm}
\subsection{Pilot 1: Business context}

\subsubsection*{Context} The first pilot study involved 
an Italian company headquartered in Milan that specializes in digital transformation services, including digital marketing and data analytics. Operating on an international scale, the company employs over 1,000 professionals. The pilot study was conducted in Milan at the company's premises on July 2, 2024, and spanned half a day.

\subsubsection*{Participants} 
The study involved 13 participants with different backgrounds and from three distinct nationalities (Italy, East Europe, Asia). They were from various departments (e.g. data science, human-computer interaction and management) and had different roles within the company, including project managers and programmers. Thus, the group was not homogeneous, and included both horizontal and vertical working relationships.. All participants understood Italian and English, so the activities were primarily conducted in Italian, with English used when participants faced difficulties expressing themselves or understanding in Italian. 

\subsubsection*{Scenario} 
We collaborated with the company to define an application scenario of potential interest for both parts. We agreed to select a real application that was actively being developed by the company, namely a virtual assistant leveraging AI to consult the company's internal knowledge base. The application was envisioned for employees to request details about policies, projects, and internal teams using natural language, with data sourced from a centralized repository of company documents. At the time of the pilot, the application was in testing, with 11 management members trialing it for one month. The company planned a full roll-out if the trial proved successful. Only a few pilot participants had prior access, as they were directly involved in its development.
The scenario involved two specific tasks: $(i)$ identifying relevant risks associated with the development and deployment of the tool,
and $(ii)$ formulating recommendations to mitigate these risks.
In addition to serving research purposes, the exercise aimed to explore potential recommendations that could inform company policies and guide future steps in the application’s development.

\vspace{-0.2cm}
\subsubsection*{Outcomes}
\begin{itemize}[leftmargin=*]
\item 
\texttt{Scenario-specific results.} 
 The concerns expressed by participants ranged from privacy and data protection to transparency. The identified risks primarily involved the quality of  the input data and the output generated by the virtual assistant (Fig~\ref{fig:samples} 1.1.),  
  as well as
  potential unintended consequences,  such as the impact on employees' autonomy (Fig~\ref{fig:samples} 1.2.), 
  knowledge-sharing 
  (Fig~\ref{fig:samples} 1.3.) 
  and collaboration (Fig~\ref{fig:samples} 1.4.). 
  Regarding recommendations, there was strong consensus on the need to establish a multidisciplinary committee to monitor the tool's quality, though participants expressed varying interpretations of what impacts quality (data,  operation, communication, feedback management, troubleshooting and releases).
  A richer list of collected risks and recommendations is in 
  Appendix~\ref{app:pilot1}.1.  
\item 
\texttt{Participant-specific results.}
  Participants expressed a genuine curiosity about the idea of carrying out moral exercises in a business context. 
  They approached the activity seriously demonstrating a strong interest in critically examining the scenario, likely driven by the fact that it is a real application with a direct 
  impact on their work.
  They remained focused and engaged throughout the activities and the intermediate pause many engaged in an exchange of views about the working methodology.
  A review of the worksheets confirms significant involvement: while notes varied in conciseness and completeness, all participants attempted to record others' contributions, with someone also highlighting and commenting parts of exercises' instructions.
  In several cases, the style and nuances of participants’ descriptions of risks and recommendations reflected a sense of empathy and \emph{emotional investment}, covering a range of emotions that included fear, 
   distrust, 
  hope 
  and excitement (Fig~\ref{fig:samples} 1.5-1.8.). 
  Responses also suggested participants' \emph{moral commitment} in developing 
  technology guided by values (Fig~\ref{fig:samples} 1.9.-1.11.) 
  and attetion to AI-human integration  (Fig~\ref{fig:samples} 1.12.) 
  See Appendix~\ref{app:pilot1}.2 for a more comprehensive sample of annotations.
\item  
 \texttt{Exercises-specific results.} According to the follow-up questionnaire, during the exercise participants felt highly comfortable (mean score 4.8 over 5) and stimulated in ethical reflection (mean 4.2). Interestingly, they fully agreed that ``stimulating the practice of one's moral abilities is useful in their profession'' (mean 5). Both the activity and the emerged recommendations has been judged very useful (mean 4.8).
  Participants appreciated the practical approach of the exercise as they expected a more speculative discussion on general ethical principles. 
  Regarding the challenging aspects, participants remarked limited time for insights elaboration in step 3, uncertainty about how to structure their thoughts at the beginning of the exercise, and difficulty in active listening and selecting only one topic. Regarding the positive aspects, participants appreciated the exercise’s emphasis on deep listening, reflection, and careful attention to words. They valued the opportunity to engage with others' ideas, take time to identify key insights, and express themselves freely. The structured approach to active listening and group facilitation was seen as a valuable support also for stakeholder consultations and extracting diverse opinions. Someone noted that exercise's focus is on ``democratizing ideas'' rather than synthesis. 
  With regard to the methodology, we observed further practical challenges with manual writing: while  almost every worksheet was complete in reporting others' entries, 
  many people condensed what they heard from others using keywords and summaries (see Section~\ref{sec:discussion} for a discussion). Another practical challenge was 
  the multilingual approach. We provided instructions in both Italian and English, encouraging participants to express themselves in their preferred language before translating into English. Although everyone was proficient in English, we observed a tendency to switch to Italian when sharing personal views. This suggests that a \emph{lingua franca} like English may not always be the most effective medium for engaging participants' moral dispositions.
\end{itemize}

\begin{figure}[t]
\raggedright
{\small 
\begin{center}
    {\bf Pilot 1 - Extracts from worksheets}:
\end{center}

Scenario-specific results:\\
\begin{enumerate}[label=1.\arabic*.,leftmargin=*,align=left] 
\it
 \item Different versions of files as PPT, PDF, on the same subject, could lead to conflicting or incorrect information, or simply be out of date
\item Risk of losing the uniqueness of one’s own thinking or intuition to carry on a project.
\item I hope it will not ‘cannibalize’ the depth of knowledge and study that comes from placing humans, teams, relationships, and the synergy of minds at the center of the creative process, as well as the process of understanding context and ideas.
\item People may become less inclined to talk to each other and to seek confirmation of the acquired information.
\end{enumerate}
Participant-specific results
\begin{enumerate}[label=1.\arabic*.,leftmargin=*,align=left] \setcounter{enumi}{4}
\item \emph{To me, this is something that pushes the limits of the known and may cause fear and repulsion.} 
\item \emph{There is a risk that users will lack trust in the tool, becoming suspicious and skeptical.} 
\item \emph{I hope this virtual assistant will be widely adopted and that the company recognizes how complex it is to develop something like this.} 
\item \emph{This project is my first big project so my motivation is to share this passion with others.} 
\item \emph{Technology that HELPS work and SUPPORTS people in their everyday work.}  
\item \emph{What about a corporate code of ethics for the programmer?} 
\item \emph{How inclusive it is? (ethnic, religion,culture, gender, nationality)}
\item \emph{Where can it go from here? [...] vision of how it will change the way we work with AI.}
\end{enumerate}

%


\begin{center}
    {\bf Pilot 3 - Extracts form worksheets:}
\end{center}
\begin{enumerate}[label=3.\arabic*.,leftmargin=*,align=left]
\item \emph{Data must not be shared with anybody else}  
\item \emph{the tool shouldn't have access to students' data without explicit permission.} 
\item benefits: \emph{"Reduce the pressure of finding a job", "save time", "allow for freer expression"}; concerns:\emph{"reduce opportunities to make friends", "limit real human connections"} 
\item \emph{"it will help for new students to get help without asking"} and similarly \emph{"sometimes I have questions which is difficult to ask"}
\item \emph{the chat-bot could be a useful helper for student who don't speak the language well or who are not familiar with the administration} 
\end{enumerate}
\caption{Samples of results of pilot studies}
\label{fig:samples}
}

\end{figure}

\vspace{-0.3cm}
\subsection{Pilot 2: Research context}

\subsubsection*{Context}
The second pilot study was conducted as part of an interactive session at an international workshop on AI ethics. The proposed activity was selected through a peer-reviewed process and took place in July 2024 in Germany.
As the session was part of a larger event, we had a limited time slot of 90 minutes to carry out the exercise. For this reason, we included only the first and second steps, omitting the elaboration of insights and using simplified worksheets that focused solely on recommendations, excluding risk assessment. Additionally, given the research-oriented setting, we took the opportunity to introduce the broader concept of moral exercises and allocated time for concluding remarks. 
Although we did not distribute a questionnaire, the venue provided multiple opportunities to gather spontaneous feedback after the session.
\vspace{-0.2cm}
\subsubsection*{Participants}
The study involved 20 participants from diverse backgrounds (computer science, design, law, social science, and philosophy)
and professionals from both academia and industry. All participants were 
connected by a shared interest in the workshop’s topics, such as algorithmic fairness, and an effective commitment to multidisciplinary dialogue.
\vspace{-0.2cm}
\subsubsection*{Scenario} 
The selected scenario focused on a fictional university implementing an AI-based tutoring application to enhance student learning by offering immediate, personalized, and interactive academic support outside traditional classroom hours. The scenario specified the technologies used (i.e., pre-trained large language models), the developmental stage, and key functionalities. 
Participants were invited to take on the role of stakeholders 
and reflect on the key values at stake, 
proposing recommendations to address potential risks associated the adoption of the AI tutoring system. Given the nature of the venue, we shifted the focus from the scenarios to the exercise itself, clarifying to the audience that our primary interest was in gathering direct or indirect feedback on the methodology rather than on the AI tool. 
\vspace{-0.2cm}
\subsubsection*{Outcomes}

\begin{itemize}[leftmargin=*]
  \item 
\texttt{Scenario-specific results.}
  Regarding the AI application, the most frequently mentioned concerns included accessibility, autonomy, privacy, voluntary use, healthy usage, non-discrimination, and creativity. A recurring recommendation was the active involvement of stakeholders, particularly the most vulnerable groups, such as students with disabilities. Participants also emphasized the importance of establishing a student-teacher committee responsible for conducting regular assessments of the system. Additional recommendations emphasized \emph{``encouraging the collaborative use of the tool''}, \emph{``ensuring that it complements rather than replaces traditional schooling''}, and \emph{``promoting healthy study habits.''}
 \item 
\texttt{Participant-specific results.} 
  The participants showed attention, professional commitment and collaboration. During the reading rounds, they carefully took turns, often looking at others to ensure everyone had sufficient time to complete their annotations. The content recorded in the worksheets was detailed and insightful. However, when lengthy sentences were read aloud to the group, many participants synthesized multiple aspects rather than strictly adhering to the original wording. 
  We noticed a tendency to use standardized terminology commonly found in AI ethics literature, such as 'checking for bias,' 'diverse stakeholders,' and 'human in the loop.' This may indicate a more abstract 
  perspective on the issue, reflecting impersonal beliefs and only partial engagement with the specific problem under discussion. The topic may have seemed more theoretical to them, unlike in Pilot 1, where the tool had direct relevance to their personal or professional lives. 
\item  
 \texttt{Exercises-specific results.}
  We received overall positive feedback, and the fact that several participants (4–5) expressed interest in replicating the activity in their own contexts suggests that the exercise successfully encouraged moral reflection. Notably, Pilot 3  is as a direct follow-up to this event. As in Pilot 1, participants appreciated the structured nature of the activity and its emphasis on active listening, which ensured that all voices were heard. However, some found it unclear whether the final task was intended to focus on the objective frequency of recurring themes or their subjective interpretation. One participant suggested that for relatively simple issues, the activity might be overly complex, and that simpler mechanisms, such as preference aggregation, could be more suitable. Additionally, there were concerns that marginal voices might be overlooked in the final synthesis.
  We believe these concerns are partly due to modifications in the exercise format due to time constraints. Specifically, we had to omit the step 3 (cf. Fig~\ref{fig:ex}), which  included a group discussion aimed at ranking recommendations by considering both the most frequently mentioned themes and the most isolated perspectives that the group deemed important to retain. Furthermore, unlike in Pilot 1, participants in this session had very limited time for their initial personal analysis in step 2, and they missed out the final review and elaboration of the short scripts included in the omitted step 3.
  The absence of these components likely influenced participants' perception of the exercise’s purpose and effectiveness.
\end{itemize}

\vspace{-0.4cm}
\subsection{Pilot 3: Student education}

\subsubsection*{Context}
The third pilot study took place on December 13, 2024, at a German university and involved higher education students. The activity was conducted as part of an English-taught course on Trustworthy AI, open to students from any degree program.
To prepare students  we held a preliminary online seminar covering moral exercises, the underlying ethical framework, and their relevance to human oversight of AI systems. The moral exercises activity was then conducted in person a few days later during a scheduled class, lasting one and a half hours. 
Similar to the second pilot, this study omitted the Insights Elaboration phase, focusing instead on Problem Exploration. The prior online seminar allowed us to streamline the introduction phase, enabling a more efficient start to the activity. As a result, we had some time at the end for an informal exchange of feedback. We used the same simplified worksheets as in Pilot~2.

\subsubsection*{Participants}
The activity involved 15 participants from computing-related fields, including computer science, digital humanities, and interdisciplinary studies combining computer science and natural sciences. In addition to German students, the group included participants from various countries, such as South Korea and Norway.
\vspace{-0.2cm}
\subsubsection*{Scenario}
The scenario closely resembled that of Pilot Study 1, but this time, participants took on the role of individuals who might be directly affected by the AI system—specifically, students. The scenario centered on an AI-based tutoring system designed in the form of a chat-bot to offer academic guidance, career advice, emotional support, and practical tools for time and task management. Participants were asked to reflect on potential risks associated with the adoption of the AI tutor and to formulate recommendations to be shared with university administrators.
\vspace{-0.2cm}
\subsubsection*{Outcomes}

\begin{itemize}[leftmargin=*]
  \item \texttt{Scenario-specific results.}
  Participants largely agreed on recommending privacy, data protection (Fig~\ref{fig:samples} 3.1.) 
  and transparency (Fig~\ref{fig:samples} 3.2.). 
  Another key recommendation regarded human direction since the chat-bot should be only a \emph{``first point of contact''} not a replacement for human.
  \item \texttt{Participant-specific results.} 
  All participants actively engaged with the scenario, demonstrating focus and a collaborative approach throughout the activity. While this level of engagement might be expected given the educational setting, the review of the worksheets provided clear evidence of their commitment. Most participants carefully documented contributions from all group members, utilizing the entire space available for personal notes.
  Many expressed their opinions in the first person, using phrases such as \emph{``I feel...''} and \emph{``I think...''}. Their reflections conveyed both hopes and concerns. 
  Some participants were optimistic about the AI tool’s potential, 
  others voiced concerns (Fig~\ref{fig:samples} 3.3-3.5.).
  Participants' notes also demonstrated a strong ethical awareness of potential risks, including job displacement, widening social inequalities, and increased energy consumption. 

\item \texttt{Exercises-specific results.}
 After the exercise, students provided live feedback, expressing appreciation for the balanced participation, as everyone had the opportunity to contribute. They also observed that the group discussion led to different recommendations compared to individual preferences. For instance, one participant mentioned that if working alone, they would have prioritized sustainability, whereas the group collectively focused on shared concerns such as data privacy.
 Reading aloud and taking notes were difficult for some students. Shy participants often read quickly or in a low voice, requiring repetition for clarity — an issue that was more common among non-native English speakers. To address this issue the moderators attempted to slow down the process or request participants to repeat their reading. Additionally, students who struggled with note-taking tended to produce lower-quality written reflections. Younger participants also required more time to grasp the process, particularly the importance of waiting for everyone to complete their writing before proceeding with the next step.
\end{itemize}

\vspace{-0.3cm}
\section{DISCUSSION}
\label{sec:discussion}

The pilot studies suggest that moral exercises can be effective in activating moral dispositions, enabling the subject to explore the interconnectedness of professional and domain-specific skills with moral reasoning and personal feelings. At the same time, by remaining close to technical and pragmatic aspects, they provide
useful concrete outcomes, as testified by the business context case, where a report of the outcomes has been used by the company to inform the subsequent development of the AI application under scrutiny.

In this section we assess the experimentation conducted in the three pilot studies. We discuss the merits and the limitations of both the general approach and the three main pillars of moral exercises put forward in Section~\ref{sec:pillars}.

\subsubsection*{General approach to ethical reasoning and deliberation}
While existing approaches, such as value-sensitive design \cite{friedman2019value}, already advocate for integrating ethical judgment into the design of socio-technical systems, moral exercises take this a step further. Participants' responses demonstrate that moral exercises can reconnect moral judgment with individuals’ personal experiences, reinforcing the idea that in moral life emotions are not separate from, or opposed to, reason\cite{vallor2016technology}. 
In this way, moral exercises provide a valuable counterbalance to the risk of reducing ethical assessment to a purely abstract and formal activity with little real-world impact \cite{hagendorff2022blind}. They offer the experience that the most fundamental source of ethical principles and values is the human agent, especially those directly involved in the AI project under evaluation. 
%
%
Placing \emph{people} at the center, 
draws attention to an approach that is process-oriented rather than solution-driven.
This challenges dominant AI frameworks, where a technocratic mindset and consumerist tendencies often reduce ethics to problem-solving and standardized terminology.
Our experience with moral exercises demonstrates that engaging moral and personal dispositions requires moving beyond the \emph{user}-centered approach commonly adopted in participatory design \cite{costanza2020design,spinuzzi2005methodology}. 
These exercises encourage participants to express themselves beyond their predefined roles (e.g., users or designers) and technical jargon.
Furthermore, unlike computational methods for preference elicitation, moral exercises highlight the intrinsic value of focused attention toward oneself and others, an aspect that cannot be easily captured through metrics or computational approaches \cite{feffer2023moral}.
On the other hand, pivoting on personal judgment instead of on procedures, is challenging. For instance, some participants wondered what ensures the ‘moral correctness’ of the recommendations arising from the exercise,
especially because the exercise lacks an introductory explanation of ’what is good or evil’. As already emphasized,
the moral exercise is not intended to instruct what is an ethical action, but to develop ethical guidance 
in collective discernment.
Moral
decisions can hardly be sorted into correct and incorrect, but a quality decision process can be more effectively
characterized. Accordingly, the pillars of moral exercises, based on the activation of practical wisdom and engaged responsibility, 
stimulate the development of the moral compass that can
guide to fair decisions.

\vspace{-0.2cm}
\subsubsection*{Personal commitment and self-examination}
We observed that organizing the exercise as an in-person meeting encouraged participants to approach moral cultivation as an integrated activity, combining both physical and non-physical aspects. This served emphasizing that moral reasoning is not purely mental process but is shaped by physical experiences, an intuition well-entrenched in various philosophical theories (e.g. the Ethics of care, Aristotelian virtue ethics) and cultural and religious traditions (Buddhism and Christianity). 
The downside is that it is difficult to express (or measure) the impact of activity on people, whereas it would be very useful to equip the AI supervisors with indicators to become aware of the progress. 
In this regard, we should investigate criteria for self-assessment consulting psychological or pedagogic scholarship. 

The pilot studies also showed that special care must be placed in the choice of the scenario of the exercise, which needs to be close to the experience of the participants, whose effective engagement is crucial.
This opens the issues of how to
create scenarios engaging people who may lack a shared experience, or who only meet on a single occasion just to experiment with this methodology.
At the same time, the experimentation acknowledged
the importance of the training of moderators of the exercises.

\vspace{-0.2cm}
\subsubsection*{Relational understanding}
One of the most interesting finding is the effectiveness of reading rounds in ensuring that all voices are heard. This method was largely appreciated by participants but may pose challenges in contexts with asymmetric power dynamics. In this regard, informal feedback from Pilot Study 1 revealed contrasting perspectives. Some participants viewed the approach as a way to equalize voices, while others expressed concerns about the presence of hierarchical roles. While we recognize that in workplace settings this is a sensitive issue requiring context-specific evaluation, under safe conditions the exercise can foster dialogue and strengthen constructive relationships within a professional environment.
In addition, since the task was simply reading --on a voluntary basis-- the written text without commenting, all people offered their own contributions 
without anyone prevailing in or monopolizing the conversation. Mutual listening is also were focused attention was put under strain. Recording what the others read was one of the most challenging task reported by participants and evidenced by the review of worksheets. Though recording was not a goal in itself, the effort gave participant the opportunity to materially take care of the other's voice. This exercise also served as a safeguard against common tendencies that can undermine the quality of dialogue, such as focusing on formulating a response while the interlocutor is speaking.
\vspace{-0.15cm}
\subsubsection*{Methodology}
Pilot studies confirmed the value of key methodological choices. 
Manual writing required participants to engage more consciously with active listening and discernment. Additionally, it prevented unnecessary distractions that might arise, for example, when using digital tools, allowing participants to remain focused on the essential aspects of the exercise. Note that no participant suggested to move to digital supports. Moreover, our observations align with other community-driven practices of data curation\cite{mcmillan2024data}.

We also recognize the benefits of providing limited space and time (approximately 8-10 min slots) for participants' notes. This approach not only encouraged conciseness during the reading rounds but also reduced the temptation to over-edit or strive for all-encompassing opinions. We believe that embracing the imperfections and limitations of each contribution serves as tangible evidence that no single individual can provide a complete perspective on the problem - complex issues require the contributions of diverse individuals to be fully explored. Naturally, collective contributions are unlikely to produce a fully satisfactory solution. However, as observed in Pilots 2 and 3, the process of elaborating insights enabled the group to better identify points of convergence, divergence, and gaps. 

As a flexible methodology without a rigid structure or protocol \cite{Sca2023} — aligned with the rules-as-model perspective \cite{rulesDaston} — designing a moral exercise requires careful adaptation to the specific context, including its target audience, purpose, and expected outcomes. Our experimentation highlighted the importance of effective time management, balancing personal reflection with structured group discussions. Enhancing this balance would allow for deeper individual contemplation before group exchange and ensure sufficient time for finalizing insights, preserving the depth and richness of the discussions.
An open question remains whether to adapt the exercise format for shorter time slots or establish a minimum duration for effective implementation. Meaningful personal and group reflection is essential; without it, the exercise risks becoming a mere data collection exercise rather than a genuine ethical deliberation.

\vspace{-0.2cm}
\section{Conclusions}
\label{sec:conclusion}
This article has described the practical exploration of a structured activity designed to cultivate and enhance the ethical skills and moral judgment of
individuals involved in decision-making processes, particularly in the oversight of AI systems.
We called it moral exercises to emphasize that their ultimate goal is not just to inform but to transform AI actors so as to create a change in mindset \cite{borenstein2021emerging}.
We situated our approach in an ethical framework, highlighting its links with  virtue ethics \cite{vallor2016technology} and the rules-as-model approach \cite{rulesDaston}, and we underscored the main pillars of this methodology,
namely, eliciting an engaged personal disposition, enabling the relational understanding and training the technomoral wisdom.
To evaluate the feasibility and practical implications of moral exercises in a socio-technical context, we conducted three pilot studies. Beyond generating concrete technical insights related to the AI scenario under review, the exercises also produced valuable non-technical outcomes among participants, including increased engagement, interest in the methodology, enhanced moral reasoning, and greater awareness of professional skills beyond technical expertise.
These findings suggest that moral exercises can strengthen both participants' and moderators' ability to identify key issues and improve deliberation processes, contributing to more effective human oversight. Future work will explore alternative exercise formats and conduct a deeper analysis of the role and training of moderators.

\bibliographystyle{plain}
\bibliography{Biblio}

\appendix
\section{Outcomes of Pilot 1}\label{app:pilot1}

\subsection{Sample of risks and recommendations collected during the moral exercise.}

 {\small
 \begin{itemize}
     \item \emph{"Risk of losing the uniqueness of one's own thinking or intuition to carry on a project, by focusing only on the development process or on the ideas recommended by the tool." (Rischio di perdere l'unicità del proprio pensiero o sensazioni per portare avanti un progetto, considerando unicamente flussi di sviluppo o idee che il tool potrebbe
 consigliare)
 }
 \item \emph{"Risks concerning the generation of new knowledge/ideas/proposals/vision. People are less inclined to talk to each other and confront each other, to ask for confirmation of the acquired information (of correct deductions). People are less stimulated to connect, generate, shape, err."}
 \item \emph{"If it’s only used by a small group of constant users, it may create stronger bias over time."}
 \item  \emph{"Different versions of files, such as ppt, pdf, on the same subject, could lead to conflicting or incorrect information, or simply, out of date."}
 \item \emph{"Ensure accessible, non-technical, language in the preparation of presentation materials and information documents."}
 \item \emph{"Convey clear message what this tool can do and how to use, and what to expect, emphasising that this is continuous development so with the time it will get better."}
 \item \emph{"On the basis of an accurate and as complete as possible test, define and share rules of usage (what the AI assistant can and can't do, how to use answers, promoting rules,..)."}
 \item \emph{"Having a plurality of resources (even raw ones), to document the process followed in projects, the things that
 did not work and those that went well."}

 \item \emph{"Instead of developing a powerful AI GPT, we should put effort/time to build a
 consolidated and robust data source that could provide high quality data to the virtual assistant, then benefit user."}
 \item \emph{"Leave to users full control on managing their data and how to make them accessible to others."}
 \item \emph{"The output will contain links to documents used to answer?"}
 \item \emph{"If the core tech is OpenAI, does it mean we have limited control over topics (ethical)?}
 \end{itemize}
 \label{fig:report}
 }

\subsection{Sample of observations collected during the moral exercise.}

 {\small
 A number of observations had an {\bf emotional nuance} such as fear, laziness and trust: 
 \begin{itemize}
     \item \emph{"To me this is something that pushes the limits of known and that can cause fear and repulsion."
     \item "The tool is making everyone who was building it and using it more profound".} 
     \item \emph{
 "Risk feeding people's laziness: having what we need always accessible make us more efficient?" }

 \item \emph{"Risk of a lack of trust on the tool, users might be suspicious and mistrust it (guardare con diffidenza) because its purpose and objectives are not clear"}, and similarly  \emph{"to collect the doubts of those who will use the product in order to BUILD TRUST."}
 \item \emph{"I expect it to solve a superficial need and I hope it will not ‘cannibalise’ an in-depth knowledge/study 
 that goes through putting the man, the team, the relationships and the synergy of minds at the center of the
 creative process, and the process of generating understanding of context and ideas".}
 \item \emph{"People will not ask freely since they think they are monitored (e.g. I don't ask this maybe it is stupid."}
 \end{itemize}

 {\bf Personal expectations} and {\bf emotional investment} were sometimes clearly expressed, by writing notes using the first person singular:
 \begin{itemize}
 \item \emph{"There is a great desire to prove that AI, this AI, is useful. This expectation is sometimes very heavy and prevents one from thinking about risks or alternatives. In itself it is an exciting project, seeing engineers and designers collaborate is beautiful. Certainly not the easiest thing. Inside I hope that this virtual assistant will be used by many, and that the company understands how complicated it is to develop something like this."
 \item
 "A big risk is that nobody uses it. This is the thing that worries me the most. People might find it useless, and then we wouldn't have enough data to improve it."
 \item "This project is my first big project so my motivation is to share this passion with others".
 }

 \end{itemize}

 {\bf Moral aspects} often cover the impact on the working environment: 
 \begin{itemize}
 \item \emph{"The process of resource creation is not improved, [instead] information retrieval is improved, [thus] collaboration (which is what we need most) may not be improved".}
 \item \emph{"Can flatten internal knowledge by returning partial options (corresponding to the best-indexed data)".}
 \item \emph{"We have values, but what about a corporate code of ethics for the programmer?"}
 \item \emph{"Where can it go from here? [...] vision of how it will change the way we work with AI."} and similarly \emph{"How will it be evolving? now as a info provider, later?"} 
 \item \emph{"How inclusive it is? (ethnic, religion,culture, gender, nationality)"}

 \item \emph{"If everything is standardised we will not go wrong but by making mistakes we may come across unexpected opportunities and
 options available only because they are outside our control and knowledge."}
 But also \emph{"More knowledge for all, faster growth, more uniformity"},  and \emph{"To reduce search time and streamline the process of 'understanding', 'study' and ’discovery’."}
 \item \emph{"This tool was born from a COMMON NEED (underlined), which emerged from the needs of those who work here. I like the
 the fact that it is not just a top-down idea".}
 \item \emph{"The idea of technology that HELPS work and SUPPORTS people in their everyday work and that can bring order to chaos".}
 \end{itemize}
 }

\subsection{Extracts from the feedback forms}\mbox{ }

 {\small
  What did you find most difficult?
  \emph{
 \begin{itemize}
  \item    
 Not having enough time for group work. This made it difficult to have an open discussion with other group members. 
 \item
 Probably most difficult was the beginning, since I was confused in what way to shape my thoughts.
 \item
 Active listening and 
 selecting only one topic leaving the others behind; it was difficult to choose the most important one.
 \item
 Choosing one statement among many very similar ones
 expressed by the participants.
 \end{itemize}
  }

 What did you appreciate most?
 \emph{
 \begin{itemize}
 \item 
 The attention to words, to intuition and to deep listening. 
 \item
 Mostly I appreciated listening to the idea's of others and dedicating time to pick the best idea among ideas of others.
 \item
 Active listening,
 the mode that was adopted that allowed everyone to express themselves by giving time to reason and share their thoughts and ideas.
  \item
 A type of facilitation of group work dynamics different from that usually adopted.
 \item
 The stimulus to reflect and listen.
 \end{itemize}
 }

 Free comments and suggested changes:
 \begin{itemize}
 \item \emph{A small suggestion: organise all the worksheets into folders that are distributed at the beginning. Managing the various sheets was time-consuming and it was also easy to get confused when retrieving the required information (moreover, a rigid folder, with a pen, would help to write more comfortably). 
 Apart from that, it was a great experience that I hope to repeat!}
 \item \emph{ I wouldn't change anything.}
 \item \emph{Add an aperitif at the end.}
 \item \emph{You were great, thank you for amazing activity. It really made me reflect and widen my horizons! Keep going, and looking forward meeting you in the future :)  }
 \item \emph{It was a stimulating, enjoyable and engaging activity, it had been a long time since I had been involved in something so interesting other than the usual talks or conferences.}
 \item \emph{It might be interesting to anonymise the answers and the selection of answers.}
 \item \emph{I would investigate more the how than the what (which already belongs to the way we work as designers). The difficulty we have is often to bring clients and our colleagues on board about ethical ways of doing things because they are often at odds with the way we do business.} After this suggested change, possibly the same participant  posted the following free comment: \emph{
  The name Moral Exercises created an expectation in me that I would finally participate in a unfiltered business conversation
 where I would question how we do things, rather than the
 things we do. Maybe it's a professional distortion, but in business, I think my
 misanderstanding may be common. However, I appreciate that the way we worked was been focused on listening and very democratic.
 }
 \end{itemize}
 }

\section{Sample of worksheets}
\label{app:sheets}

\begin{figure}[h]
\begin{tabular}{|c|c|}
\hline
\includegraphics[width=7cm]{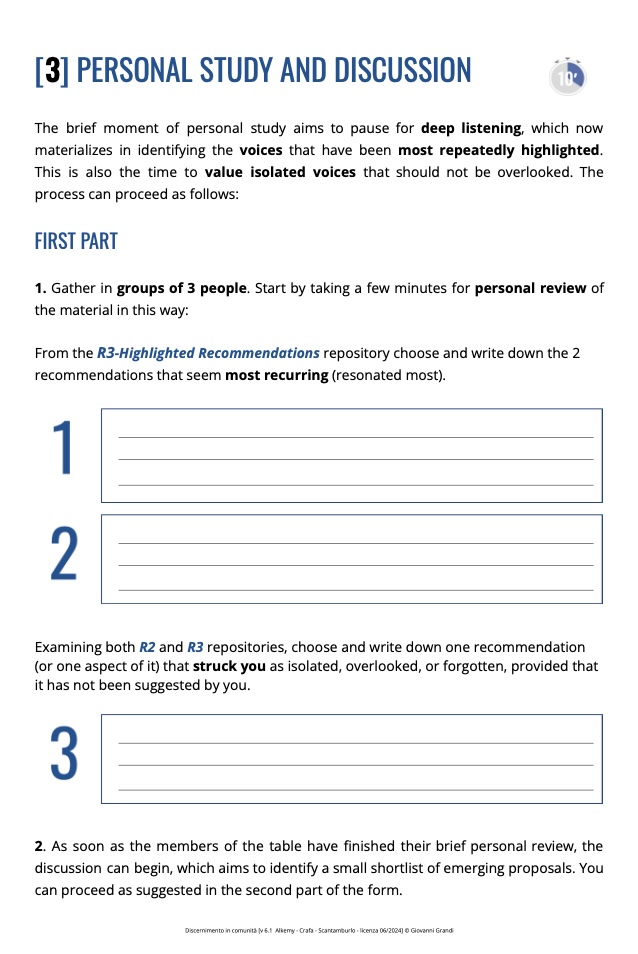}
&
\includegraphics[width=7cm]{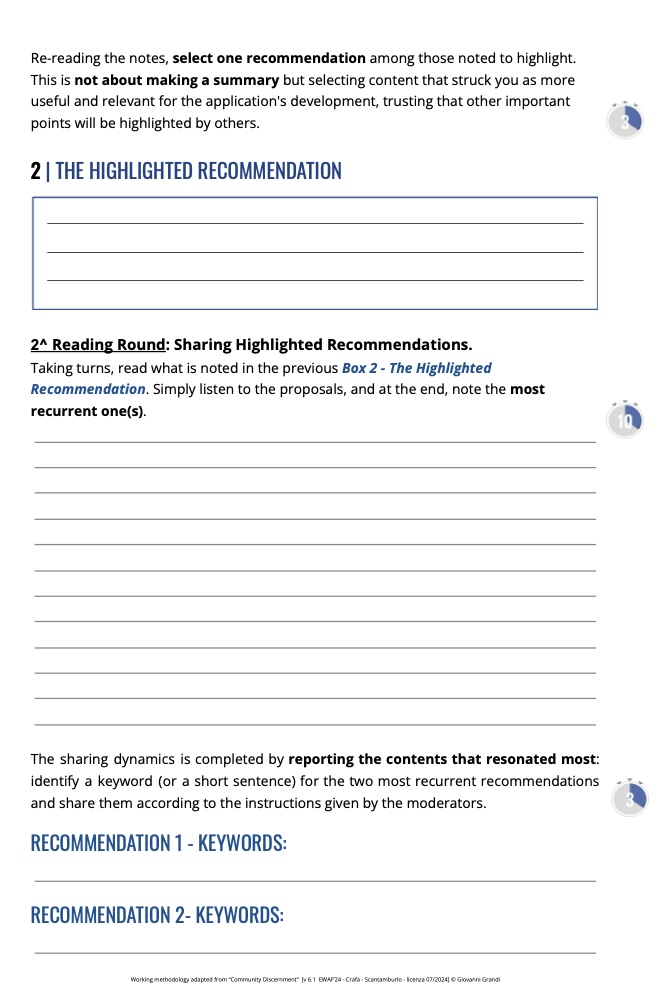}\\ 
\hline
\end{tabular}
\caption{Sample form Pilot 1's worksheet (left) and form Pilot 2 and 3's worksheet (right).}
\end{figure}

 \end{document}